\documentclass[structabstract]{aa}  
\usepackage{graphicx}
\usepackage{amsmath}
\usepackage{txfonts}
\usepackage{natbib}
\bibpunct{(}{)}{;}{a}{}{,} 
\begin{document}
   \title{Heavy-ion Acceleration and Self-generated Waves in Coronal Shocks}


   \author{M. Battarbee
          \inst{1}
          \and
          T. Laitinen\inst{2}
          \and
          R. Vainio\inst{3}
          }

   \institute{Department of Physics and Astronomy,
              University of Turku, Finland \\
              \email{markus.battarbee@utu.fi}
         \and
             Jeremiah Horrocks Institute for Astrophysics and 
Supercomputing, University of Central Lancashire, United Kingdom
         \and
             Department of Physics, University of Helsinki, Finland
             }

   \date{Received June 17, 2011; accepted September 26, 2011}

 
  \abstract
   {Acceleration in coronal mass ejection driven shocks is currently considered the primary source of large solar energetic particle events.}
   {The solar wind, which feeds shock-accelerated particles, includes numerous ion populations, which offer much insight into acceleration processes. We present first simulations of shock-accelerated minor ions, in order to explore trapping dynamics and acceleration timescales in detail.}
   {We have simulated diffusive shock acceleration of minor ions ($^3$He$^{2+}$, $^4$He$^{2+}$, $^{16}$O$^{6+}$ and $^{56}$Fe$^{14+}$) and protons using a Monte Carlo method, where self-generated Alfv\'enic turbulence allows for repeated shock crossings and acceleration to high energies.}
   {We present the effect of minor ions on wave generation, especially at low wavenumbers, and show that it is significant. We find that maximum ion energy is determined by the competing effects of particle escape due to focusing in an expanding flux tube and trapping due to the amplified turbulence. We show the dependence of cut-off energy on the particle charge to mass ratio to be approximately $(Q/A)^{1.5}$.}
   {We suggest that understanding the acceleration of minor ions at coronal shocks requires simulations which allow us to explore trapping dynamics and acceleration timescales in detail, including evolution of the turbulent trapping boundary. We conclude that steady-state models do not adequately describe the acceleration of heavy ions in coronal shocks.}


   \maketitle
%

\section{Introduction}

Particle acceleration by coronal and interplanetary shocks driven by coronal mass ejections (CMEs) is widely accepted as the primary source of strong solar energetic particle (SEP) events. Particles scatter off plasma waves, cross the shock front repeatedly and gain energy on each crossing. In large events, turbulence sufficient for extended trapping can be generated by the streaming of the accelerated particles themselves, as plasma waves in the upstream are amplified by scattering particles. This diffusive shock acceleration mechanism, as presented by, e.g., \citet{1978MNRAS.182..147B}, has recently been studied quantitatively as a model of the acceleration of SEPs by, e.g., \citet{2005ApJS..158...38L}, \citet{2007ApJ...658..622V, 2008JASTP..70..467V} 
and \citet{2008ApJ...686L.123N}. 
Other noteworthy studies of shock-accelerated SEP events include, e.g., \citet{1999GeoRL..26.2145N}, \citet{2005ApJ...625..474T}, \cite{2006ApJ...646.1319T} and \citet{2007ApJ...662L.127S, 2009A&A...507L..21S, 2009ApJS..181..183S}.

Although the most abundant ion species in the solar wind, proton, is likely to dominate wave generation in shocks, minor ions are scattered off the same turbulence and accelerated as well. If the turbulence is proton-generated, the maximum rigidity obtained by the protons determines a low-wavenumber cutoff in the spectrum of plasma waves. It has been suggested that this would prevent any ions from being accelerated beyond the same rigidity in a quasi-parallel shock wave (e.g., \citet{2007SSRv..130..255Z}). This would then lead to a dependence of the maximum (non-relativistic) energy (per nucleon) of the ions on the charge to mass ratio of the form $(Q/A)^2$.

In this paper, we present the first simulations of particle acceleration in self-generated waves with minor ions included as particles contributing to the wave generation process. Instead of using a full spectrum of ion species in the solar wind, we limit ourselves to selected interesting populations, namely $^3$He$^{2+}$, $^4$He$^{2+}$, $^{16}$O$^{6+}$ and $^{56}$Fe$^{14+}$. By simulating a propagating coronal shock for an extended period of time, we investigate the spectra and maximum attained energy of each particle population, as well as gauge the effect each population has on wave generation at different wavenumbers.


\section{Model}

In our simulation model, we numerically solve the 
particle transport equation through propagating representative particles using the guiding center approximation. We approximate the quasi-linear theory by employing a pitch-angle independent resonance condition
\begin{equation}
 f_\mathrm{res}=f_\mathrm{ci}\frac{u_\mathrm{sw}+v_\mathrm{A}}{\gamma v} \label{eq:resonance}
\end{equation}
where $u_\mathrm{sw}$ is the solar wind speed, $v_\mathrm{A}$ is the Alfv\'en speed, $v$ is the particle speed, $\gamma$ is the Lorentz factor, $f_\mathrm{ci} = (1/2\pi) q_i B/m_i c$ is the ion cyclotron frequency, $B$ is the magnetic flux density and $q_i = Q e$ and $m_i = A m_\mathrm{p}$ are the charge and mass of the ion in question, and $e$ and $m_\mathrm{p}$ are the charge and mass of a proton. We scatter representative particles isotropically off plasma waves with scattering frequency 
\begin{equation}
 \nu = \pi^2 f_\mathrm{ci} \frac{f_\mathrm{res}P(f_\mathrm{res})}{B^2}\label{eq:scattfreq}
\end{equation}
where $P(f)$ is the wave power at frequency $f$. The initial wave power is scaled to give an ambient 100 keV proton mean free path of $\lambda_0=1~R_\odot$ at $r_0 = 1.5~R_\odot$, where $R_\odot$ is the solar radius.

Once particles have been swept up by the shock, they are propagated using a Monte Carlo method, focused along the mean magnetic field due to adiabatic invariance and traced in a superradially expanding flux tube with the solar wind speed inferred from mass conservation. For additional details of our simulation model, we refer the reader to  \citet{2007ApJ...658..622V}, \citet{2008JASTP..70..467V} and \citet{2010AIPC.1216...84B}.

\subsection{Particle-shock interactions}\label{Particle-shock interactions}

As particles encounter the propagating parallel step-profile shock, they scatter off the compressed plasma and may return upstream with a momentum boost. The plasma compression ratio is solved from the Rankine-Hugoniot jump conditions. The analytical return probability from a shock encounter for an isotropic particle population is given as
\begin{equation}
 P_\mathrm{ret} = \left( \frac{v_\mathrm{w}-u_\mathrm{2}}{v_\mathrm{w}+u_\mathrm{2}}\right)^{2} \label{eq:returnprob},
\end{equation}
where $v_\mathrm{w}$ is the speed of the particle in the downstream plasma frame and $u_\mathrm{2}$ the downstream plasma speed in the shock frame. However, the downstream-transferred population is no longer isotropic, unless $v_\mathrm{w}\gg u_2$. Thus, following \citet{2000ApJ...528.1015V}, we propagate and scatter particles in the downstream plasma frame up to a distance of $2\lambda$ behind the shock, where $\lambda$ is the particle mean free path. At this distance, the particle has encountered enough scatterings to warrant the assumption of isotropy and its representative weight is multiplied by the
isotropic return probability $P_\mathrm{ret}$, sending it back towards the upstream from the distance of $2\lambda$ with a randomized shock-bound pitch-angle. Particles propagate and experience small-angle scatterings in the downstream until they either return to the shock front or their cumulative return probability drops below 0.1 \%. Simulation time is not advanced while the particles are in the downstream region. This corresponds to a situation where the downstream scattering is so intense that the mean residence time downstream is negligible compared to the time between subsequent shock encounters. Using this assumption, we do not propagate the shape of turbulence into the downstream.

\subsection{Injected populations}

In our simulation, particles swept up by a shock propagating through the solar corona are accelerated and traced up along a grid extending to a distance of $300~R_\odot$. In addition to a proton population $n_\mathrm{p}(r)$ based on the solar wind density model of \citet{2005ApJS..156..265C}, we inject fully ionized helium ($^3$He$^{2+}$ and $^4$He$^{2+}$) and partially ionized heavier elements ($^{56}$Fe$^{14+}$ and $^{16}$O$^{6+}$) according to estimated solar wind abundance values. This results in H$^+$-relative abundances for $^4$He$^{2+}$, $^{16}$O$^{6+}$, $^{56}$Fe$^{14+}$ and $^3$He$^{2+}$ of $4.0\times10^{-2}$, $8.0\times10^{-4}$, $1.0\times10^{-4}$ and $1.6\times10^{-5}$, respectively.

As a large portion of the solar wind consists of thermal particles, and our strong step-like shock, with the assumed extremely intense downstream turbulence (see \S \ref{Particle-shock interactions}), injects an unrealistically large proportion of thermal particles, we model the solar wind as consisting of two kinetic populations. The majority, 99\% of particles, represents a thermal core and is passed directly downstream. A minority, 1\% of particles, is considered a partially suprathermal halo and follows a $\kappa$-distribution \citep{2008JGRA..11306102P}. This population, which has a continuous spectrum, is encountered by the shock in our simulation. The parameter $\kappa$ receives values of $6\ldots2$ going from $1.5~R_\odot$ to $3.0~R_\odot$ respectively. It should be noted, though, that the composition of the seed population, and its dependence on distance from the sun, is not known and these parameters are arbitrary.

The average thermal speed
\begin{equation}
  w_{0}=\sqrt{2 T k_\mathrm{B} \frac{\kappa-1.5}{\kappa m_\mathrm{p}} } \label{eq:thermalspeed}
\end{equation}
is based on the radial temperature profile given in \citet{2005ApJS..156..265C}, where $T$ is the proton temperature and $k_\mathrm{B}$ is the Boltzmann constant. All minor ion populations are initialised using the same distribution function
\begin{equation}
  f(v) = \frac{ n(r) \Gamma(\kappa+1) }{ w_{0}^{3}\pi^{3/2}\kappa^{3/2}\Gamma(\kappa-1/2) }
  \left[1+ \frac{v^2}{\kappa w_{0}^{2}}\right]^{-\kappa-1} \label{eq:kappa},
\end{equation}
where $\Gamma$ denotes the gamma function.

\subsection{Wave evolution}\label{Wave evolution}
The requirement for particle trapping in front of a shock is strong enough turbulence. This can be attained by wave amplification via the scattering of energetic particles upstream of the shock. A large shock-normal velocity results in large amounts of kinetic energy deposited into accelerated particles and thus larger amounts of energy deposited into upstream waves.
Also, as particles reach higher velocities, they become resonant with waves of lower frequencies.

In steady-state upstream solar wind, the evolution equation for a normalized wave power spectrum
$\tilde{P} = (V^2 / B v_\mathrm{A}) P$ 
can be written as
\begin{equation}
\frac{\partial \tilde{P}}{\partial t} +V\frac{\partial \tilde{P}}{\partial r} = \Gamma_\mathrm{w}\tilde{P}
+\frac{\partial}{\partial f}\left(D_{ff}\frac{\partial \tilde{P}}{\partial f}\right).
\end{equation}
Here, $P(r,f,t)$ is the Alfv\'en wave power spectrum as a function of radial distance ($r$), frequency ($f$) and time ($t$), $V = u_\mathrm{sw} + v_\mathrm{A}$ is the group speed of the Alfv\'en waves and $ \Gamma_\mathrm{w}$ is the wave growth rate.
$D_{ff} = (V/r_\oplus)f^{8/3} f_\mathrm{b}^{-2/3}$ is an ad-hoc diffusion coefficient. This coefficient is chosen so that an unenhanced spectrum tends towards the form of Kolmogorov turbulence $P \propto f^{-5/3}$ at $r_\oplus=1~\mathrm{AU}$ above the breakpoint frequency $f_\mathrm{b}=1~\mathrm{mHz}$, as suggested by observations (e.g., \cite{1996A&A...316..333H}). Turbulence and diffusion magnitudes are normalized to result in an 100 kev proton having mean free paths of $1~R_\odot$ at $r=1.5~R_\odot$ and $54~R_\odot$ at $r=1~\mathrm{AU}$.

Our simulation coordinates are attached to the propagating coronal shock, allowing for good numerical accuracy in the near-shock region. Wave amplification is calculated over time intervals of $1.6~\mathrm{ms}$ at grid cell boundaries. We simulate the effect of diffusion using a Crank-Nicholson method, and advect wave power along the moving grid with a Lax-Wendroff scheme utilizing a Van Leer flux limiter.

For our simulations, we have extended the wave growth approximation of \citet{2003A&A...406..735V} to accommodate for different particle masses and charges. Energy deposited into parallel-propagating Alfv\'en waves per particle scattering can be written as $\Delta E_\mathrm{w}=-v_\mathrm{A}p\Delta\mu$, where $p=\gamma m v = \gamma A m_\mathrm{p} v$ is the particle momentum and $\Delta\mu$ is the change in particle pitch-angle. Integrating all particles of a given ion population i at a set position gives the rate of change of the wave energy density as
\begin{equation}
\frac{\mathrm{d}U_\mathrm{w}}{\mathrm{d}t}=-\int\mathrm{d^3}v~ v_\mathrm{A} \gamma A m_\mathrm{p} v \frac{\langle\Delta\mu\rangle}{\Delta t}~F({\bf r},{\bf v},t),
\end{equation}
where $F({\bf r},{\bf v},t)$ is the particle distribution function.

The pitch-angle diffusion coeffient, according to quasilinear theory, is
\begin{equation}
D_{\mu\mu}=2\pi^2 \omega_\mathrm{ci} (1-\mu^2)\frac{|k_\mathrm{r}|W(k_\mathrm{r})}{B^2}
\end{equation}
where $\omega_\mathrm{ci}=Q\omega_\mathrm{cp}/A\gamma=Q e B/A m_\mathrm{p} c\gamma$ is the angular ion cyclotron frequency, $k_\mathrm{r}$ is the resonant wavenumber and $W({\bf r},k,t)\mathrm{d}k$ is the energy density of waves propagating parallel to the mean magnetic field with wavenumber in the range from $k$ to $k+\mathrm{d}k$. This can be written as
\begin{equation}
 D_{\mu\mu} = \frac{\pi}{2}\frac{Q\omega_\mathrm{cp}}{A\gamma m_\mathrm{p} n_\mathrm{p}(r) v_\mathrm{A}^2} (1-\mu^2)|k_\mathrm{r}|W(k_\mathrm{r})
\end{equation}
which, because $\langle\Delta\mu\rangle/\Delta t = \partial D_{\mu\mu}/\partial\mu$, yields the wave growth rate
\begin{align}
 \Gamma_\mathrm{w}(k) &= \frac{1}{W} \frac{\mathrm{d}W}{\mathrm{d}t} \nonumber \\
	 &= \frac{\pi}{2} \frac{Q~\omega_\mathrm{cp}}{n_\mathrm{p}(r)~v_\mathrm{A}} \int \mathrm{d^3}v~v~(1-\mu^2)|k|\delta\left(k+\frac{\omega_\mathrm{ci}}{v\mu}\right)\frac{\partial F}{\partial \mu}.
\end{align}
As in \citet{2003A&A...406..735V}, we neglect the $\mu$-dependence of the resonance condition and replace $\delta\left(k+\omega_\mathrm{ci}/v\mu\right)$ with $(1/2)\delta\left(|k|-\omega_\mathrm{ci}/v\right)$. Now, using partial integration in $\mu$, we get
\begin{equation}
 \Gamma_\mathrm{w}(k) = \frac{\pi}{2} \frac{Q~\omega_\mathrm{cp}}{n_\mathrm{p}(r)~v_\mathrm{A}} \int \mathrm{d^3}v~v~\mu|k|\delta\left(|k|-\frac{\omega_\mathrm{ci}}{v}\right)~F
\end{equation}
which can be further represented as
\begin{equation}
 \Gamma_\mathrm{w,i}({\bf r},v_\mathrm{r},t) = \frac{\pi}{2} Q_\mathrm{i} \omega_\mathrm{cp} \frac{v_\mathrm{r} S_v({\bf r},v_\mathrm{r},t)}{n_\mathrm{p}(r) v_\mathrm{A}}
\end{equation}
where $S_v = 2\pi v^2 \int_{-1}^{+1} \mathrm{d}\mu ~v\mu F({\bf r},v_\mathrm{r},\mu,t)$ is the particle streaming per unit velocity in the frame of the Alfv\'en waves evaluated at the resonant particle speed $v_\mathrm{r}=\omega_\mathrm{ci}/|k|$. We monitor particle streaming by counting whenever a particle crosses a grid cell boundary of the shock-attached tracking grid. To correct for the difference of particle streaming between the frame travelling with the shock and the frame of the Alfv\'en waves, we use an additional weighing factor of $(v_\mathrm{w}\mu_\mathrm{w})/(v_\mathrm{s}\mu_\mathrm{s})$, where $v_\mathrm{w}$, $\mu_\mathrm{w}$, $v_\mathrm{s}$, and $\mu_\mathrm{s}$ are the particle speed and pitch-angle in the Alfv\'en wave frame and shock-attached frame, respectively.


\section{Results}

We simulate three parallel, constant velocity shocks in the corona starting from $1.5 ~R_\odot$. The shock-normal velocities $V_\mathrm{s}$ are \mbox{1250 km s$^{-1}$}, \mbox{1500 km s$^{-1}$} and \mbox{1750 km s$^{-1}$}. We use zero cross-helicity for the turbulence in the downstream, while in the upstream waves propagate away from the Sun in the plasma frame. Turbulence is tracked on a logarithmic grid reaching out to 300 $R_\odot$ in front of the shock.

\subsection{Wave turbulence and its generation}

\begin{figure*}[!t]
\centering
\includegraphics[width=8.5cm]{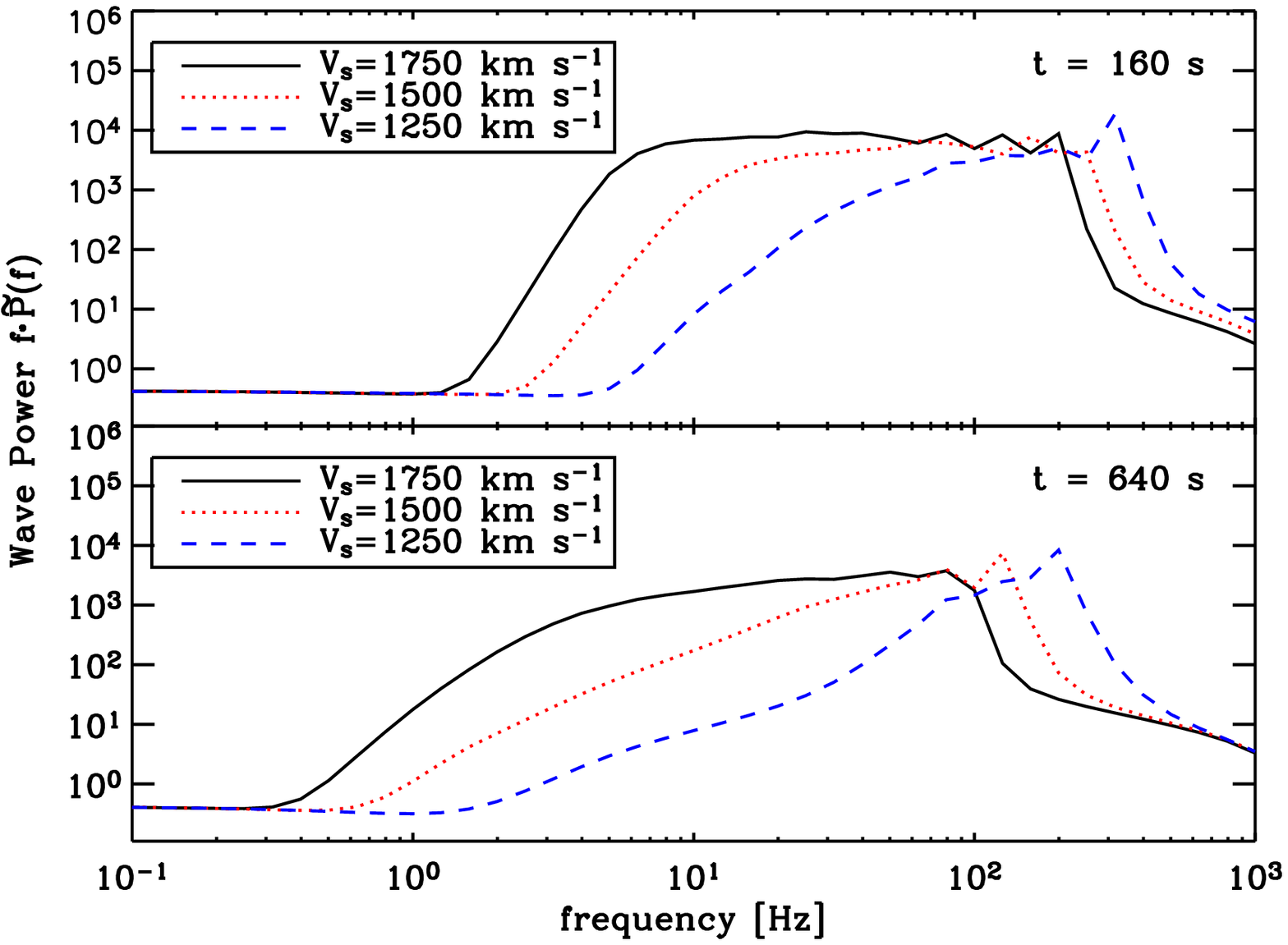}
\includegraphics[width=8.5cm]{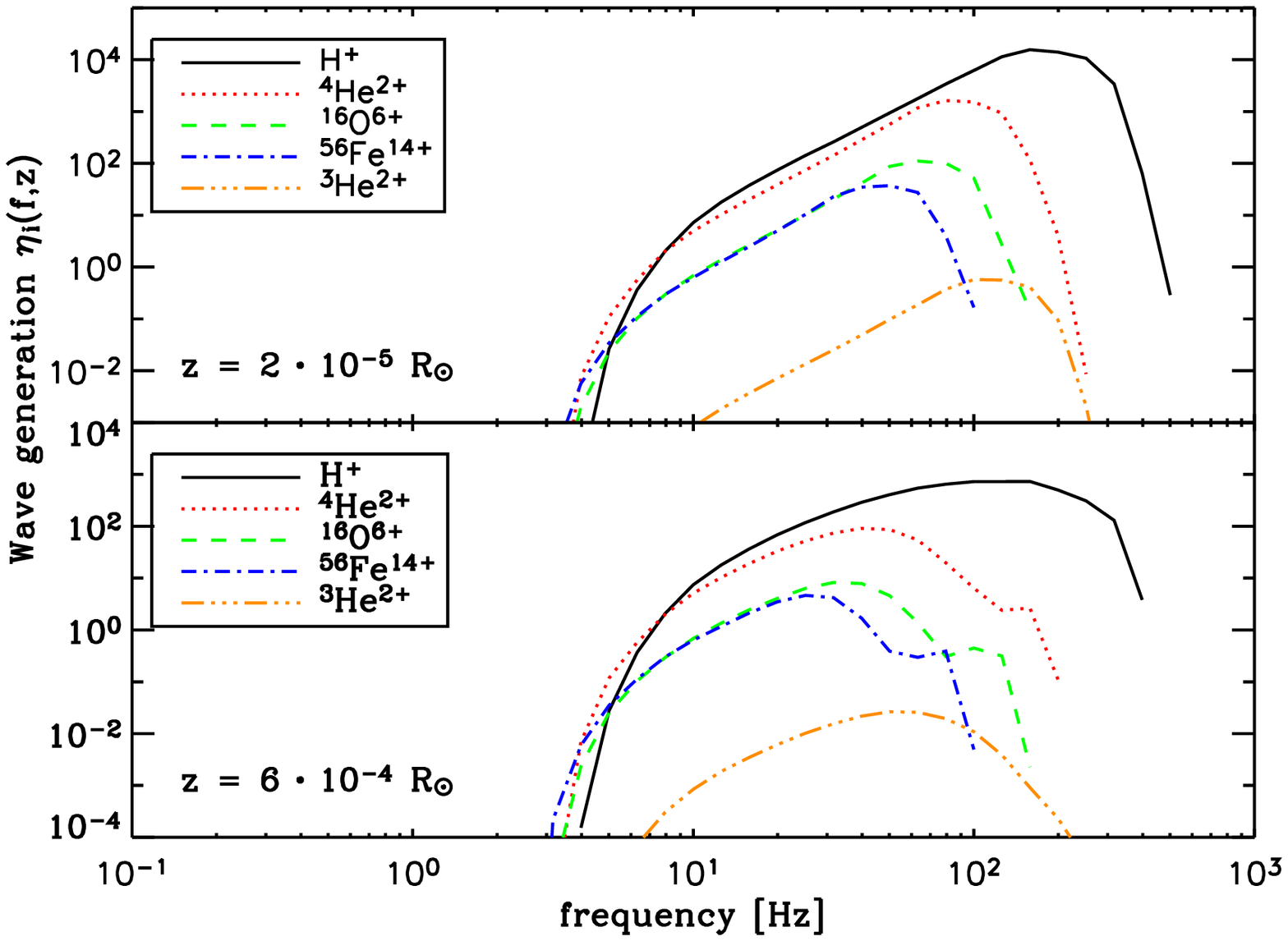}
\caption{Left: Evolution of wave power spectra. Right: Wave amplification factor $\Gamma_\mathrm{w,i}$ integrated over 640 seconds of simulation, at $2\cdot10^{-5}~R_\odot$ and $6\cdot10^{-4}~R_\odot$ from the shock, where $V_s=1500$ km s$^{-1}$}
\label{fig:wavespectra_both}%
\end{figure*}

The left panel of Fig. \ref{fig:wavespectra_both} shows the normalized wave power spectra (multiplied with the frequency $f$) in front of the shock after 160 and 640 seconds of simulation. In the right panel we show the wave amplification, $\eta_\mathrm{i}(f,z) = \int \Gamma_\mathrm{w,i}(f,r_\mathrm{shock}+z,t)~\mathrm{d}t$, for each particle population $i$, integrated over the whole simulation time. The value $z$ is the distance from the shock front and $r_\mathrm{shock}$ is the position of the shock. We display wave amplification for the \mbox{$V_\mathrm{s}=1500$ km s$^{-1}$} run in both the measurement cell nearest to the shock ($z=2\cdot10^{-5}~R_\odot$) and further out ($z=6\cdot10^{-4}~R_\odot$). Although minor ions are much less abundant than protons, they have higher charges and are resonant with lower frequencies, allowing them to generate a significant amount of turbulence close to the shock. Such dominance of minor ions has been reported by \citet{1982JGR....87.5063L} for Helium ions at low frequencies in relation to Earth's bow shock. In our simulations, close to the shock, heavier ions surpass protons in wave generation at a narrow frequency range below 7 Hz. Between 7 Hz and approximately 70 Hz, $^4\mathrm{He}^{2+}$ displays wave generation equal to $\sim50~\%$ of that of $\mathrm{H}^{+}$.

For the $V_\mathrm{s}=1250$ km s$^{-1}$ shock, $^4\mathrm{He}^{2+}$-powered wave generation is equal to $\mathrm{H}^{+}$-powered generation in the region below 100 Hz, with $^4\mathrm{He}^{2+}$ dominating below 30 Hz. As the shock-normal velocity increases, the dominance of protons on turbulence amplification increases, with the $^4\mathrm{He}^{2+}$ takeover moving to 3 Hz. In all cases, however, protons are responsible for the bulk of turbulence amplification, as abundant low-energy protons generate a great deal of turbulence above 100 Hz.

\subsection{Accelerated particle populations}

\begin{figure*}[!t]
\centering
\includegraphics[width=17cm]{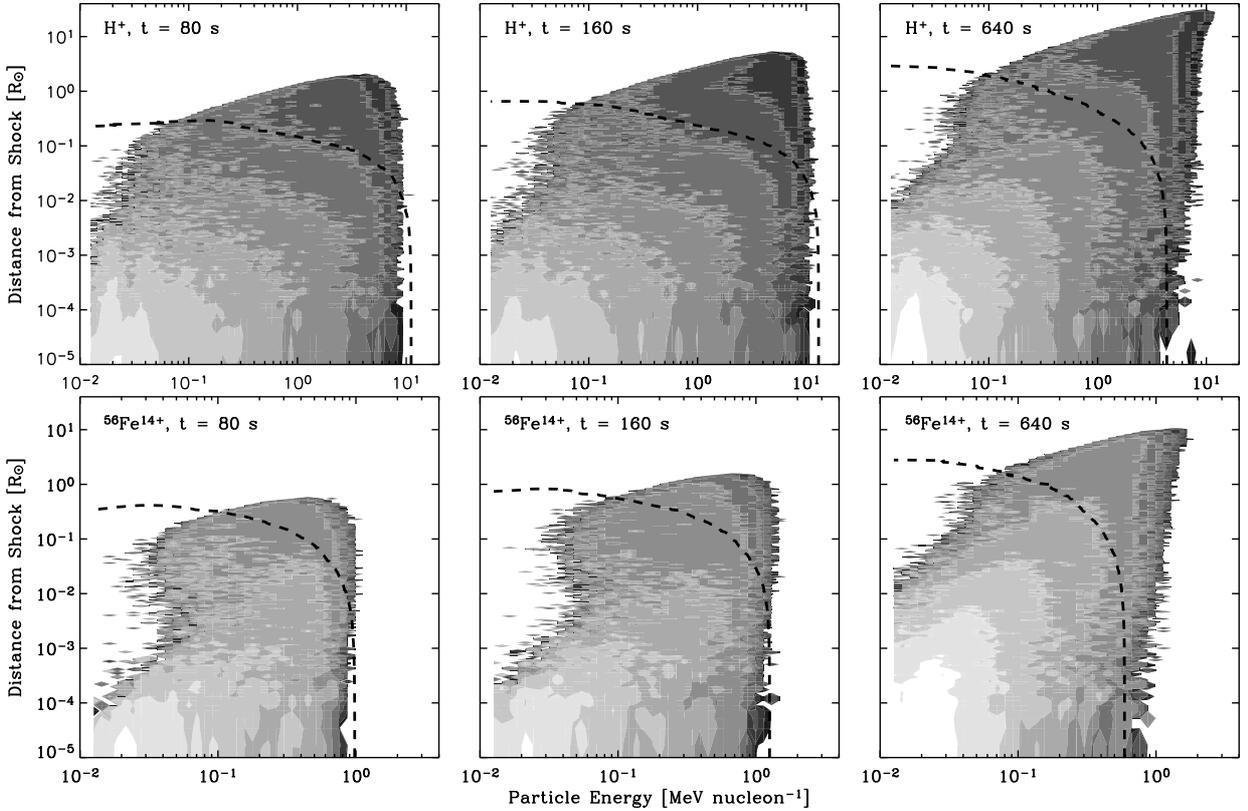}%
\caption{Particle populations for H$^{+}$ and $^{56}$Fe$^{14}$ along with the critical contour where the focusing velocity $V/L$ exceeds the shock velocity. Results are shown for the simulations where $V_\mathrm{s}=1500$ km s$^{-1}$, with colour contours at one magnitude intervals. A radial 3-cell boxcar smooth function has been applied.}
\label{fig:contour_6pics}
\end{figure*}

In Fig. \ref{fig:contour_6pics} we demonstrate the evolution of two particle populations, H$^{+}$ and $^{56}$Fe$^{14+}$. Particles within the expanding flux tube are efficiently accelerated up to a maximum energy at which the turbulence can no longer trap the particles, and instead the focusing effect of the diverging magnetic field allows them to escape.

The focused diffusion model of particle transport has been examined in detail by \citet{1996SoPh..165..205K}. Here, we start with Parker's equation, which in the fixed frame, upstream of our coronal shock reads
\begin{equation}
\frac{\partial f_{0}}{\partial t}+V\frac{\partial f_{0}}{\partial r}-\frac{p}{3}\frac{1}{\mathcal{A}}\frac{\partial}{\partial r}(\mathcal{A}V)\frac{\partial f_{0}}{\partial p}
=\frac{1}{\mathcal{A}}\frac{\partial}{\partial r}\left(\mathcal{A}D\frac{\partial f_{0}}{\partial r}\right),
\end{equation}
where $f_{0}$ is the isotropic part of the distribution function, $D=(1/3) \lambda v$ is the spatial diffusion coefficient, $\lambda=v/\nu$ is the particle mean free path and $\mathcal{A}$ is the flux-tube cross-sectional area related to the focusing length $L$ by $L^{-1}=\mathcal{A}^{-1}~\partial \mathcal{A}/\partial r$. Parker's equation can be expressed, using the linear density $n=\mathrm{d}^{2}N/\mathrm{d}r\,\mathrm{d}p=4\pi p^{2}\mathcal{A}f_{0}$, as a Fokker-Planck equation
\begin{equation}
\frac{\partial n}{\partial t}+\frac{\partial}{\partial r}\left[\left(V+\frac{D}{L}\right)n\right]
-\frac{\partial}{\partial p}\left[\frac{p}{3}\left(\frac{\partial V}{\partial r}+\frac{V}{L}\right)~ n\right]
=\frac{\partial}{\partial r}\left(D\frac{\partial n}{\partial r}\right),
\end{equation}
which shows that the effect of focusing in the particle motion is two-fold: it contributes to the advection velocity by $\dot{r}_{\mathrm{foc}}=D/L$ and to the adiabatic energy changes by $\dot{p}_{\mathrm{foc}}=-(p/3)~ V/L$. The addition to the advection velocity at large distances from the Sun is large, since there the waves have not yet grown to make $D$ small. It is clear that particles will, on average, move away from the shock in the upstream region in areas where $V+D/L>V_{\mathrm{s}}.$ This facilitates the escape of particles from the shock to the upstream and the distance where $V+D/L=V_{\mathrm{s}}$ can be regarded as the boundary of the turbulent trapping region ahead of the shock. This boundary, displayed in Fig. \ref{fig:contour_6pics} as a dashed line, outlines an escaping population further away from the shock. In addition, the energy at which the boundary intersects the shock surface is representative of the maximum energy that the particles can be accelerated to at a given time.

In the latter stages of the simulation, we find decreased wave generation due to lower particle densities and thus less swept-up particles. This, along with wave diffusion, results in the turbulent trapping boundary at the shock moving to lower energies, which causes high energy particles to escape instead of experiencing further acceleration.

\subsubsection{Spectral indices}
\label{Spectral indices}

\begin{figure*}[!t]
\centering
\includegraphics[width=8.5cm]{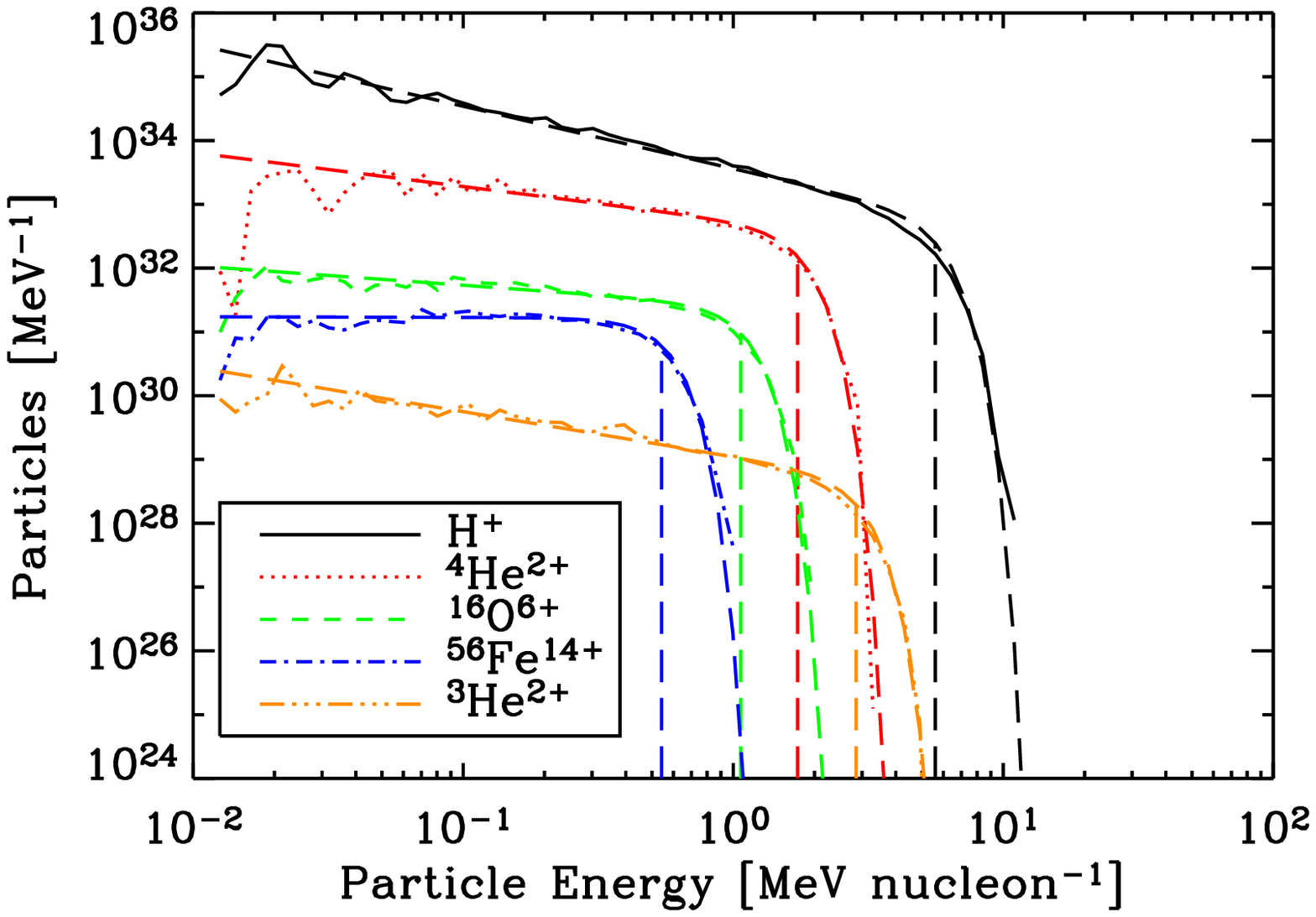}
\includegraphics[width=8.5cm]{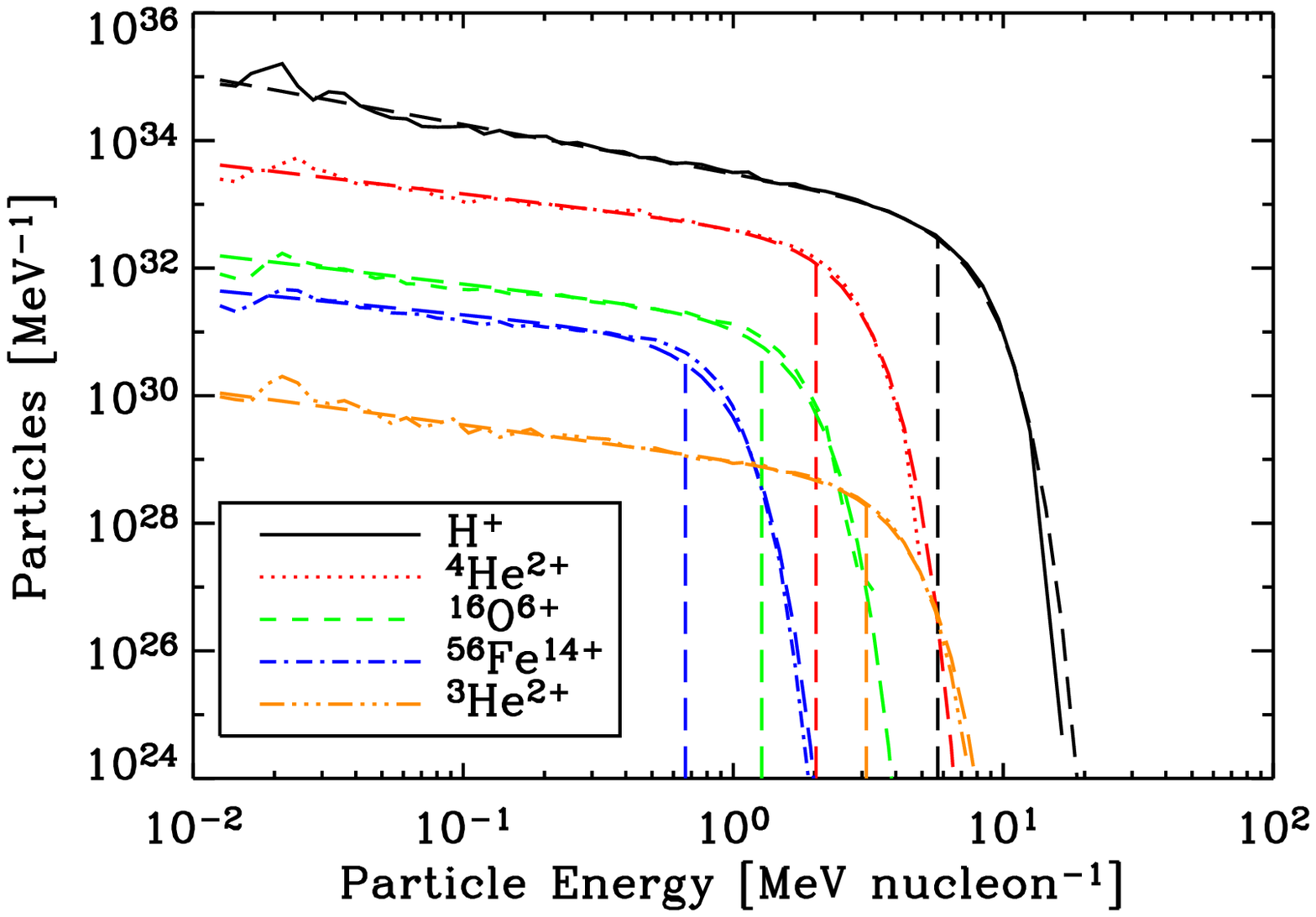}%
\caption{Particle spectra after 80 seconds (left panel) and 640 seconds (right panel) of simulation, where $V_\mathrm{s}=1500$ km s$^{-1}$. A power-law and an exponentional cut-off has been fitted to each population, ignoring the enhancement at the lowest energies.}
\label{fig:fits_1500}%
\end{figure*}

The accelerated particle populations were integrated over the whole upstream. We found the spectral index $\alpha$ for the power-law part of the particle spectrum by fitting a line to a chosen section of the log-log representation of data points. Observing the one magnitude contours in Fig. \ref{fig:contour_6pics}, we see that the spatial distributions of iron and protons differ for both the escaping population and particle populations within the turbulent trapping boundary. For iron, this results in much harder particle spectra than what the steady-state model of \citet{1978MNRAS.182..147B} suggests. Spectra along with more complete parameter fits are exemplified in Fig. \ref{fig:fits_1500}.

\begin{figure}[!t]
\centering
\resizebox{\hsize}{!}{\includegraphics{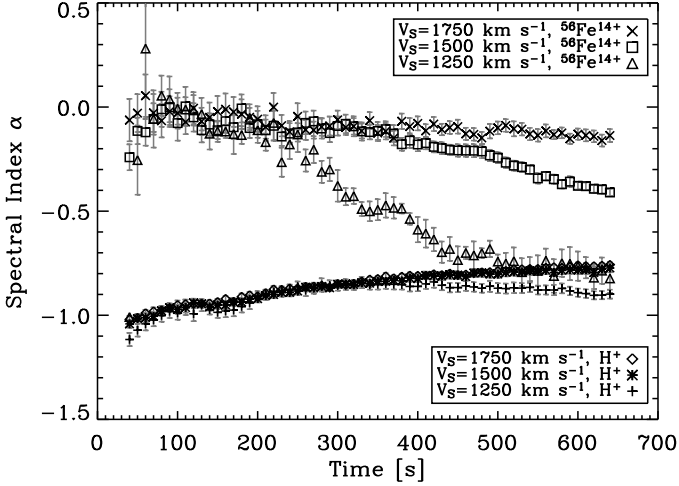}}
\caption{Temporal evolution of spectral index $\alpha$ for H$^{+}$ and $^{56}$Fe$^{14+}$ populations, as fitted to the power-law section of the particle spectra.}
\label{fig:evolution_gamma}
\end{figure}

As the simulation continues the spectrum for heavy ions softens. Figure \ref{fig:evolution_gamma} displays the evolution of the spectral indices $\alpha$ for H$^{+}$ and $^{56}$Fe$^{14+}$. Protons, being the dominant particle population, do not exhibit a softening of the spectral index, whereas the effect is exceedingly prominent in the case of iron accelerated by a $V_\mathrm{s}=1250~\mathrm{km s^{-1}}$ shock.

\subsubsection{Attained maximum energies}
\label{attained maximum energies}

\begin{figure}[!t]
\centering
\resizebox{\hsize}{!}{\includegraphics{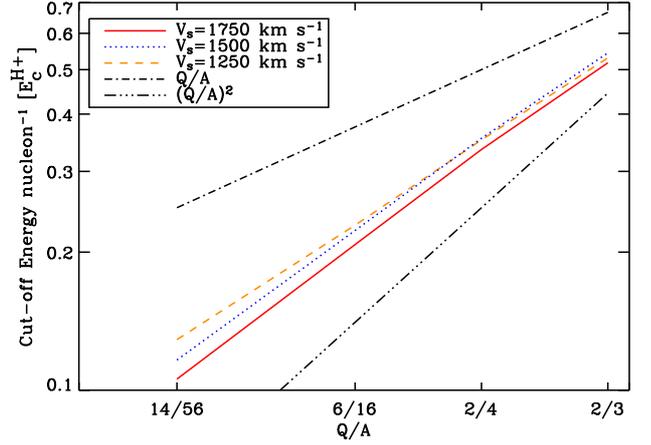}}
\caption{Ratios of particle cut-off energy to the charge/mass number as a log-log-plot.}
\label{fig:cutoff_ratios_640}
\end{figure}

When attempting to gauge the maximum energy attained by a particle population, we attempted to fit a power law with an exponential cut-off to the simulated spectrum. First, we found the spectral index $\alpha$ for the power-law part of the particle spectrum, as in \S \ref{Spectral indices}. We then used this as the basis for fitting an exponential cut-off energy $E_c$. The form used is
\begin{displaymath}
y_\mathrm{i} = C E_\mathrm{i}^{\alpha}\mathrm{e}^{-\left(\frac{E_\mathrm{i}}{E_\mathrm{c}}\right)^\epsilon}
\end{displaymath}
where $C$ is a fitted constant and $\epsilon$ is chosen to fit the sharpness of the cut-off. In our work, we used values of $\epsilon=4\ldots2.5$, with the value decreasing over simulation time. Figure \ref{fig:cutoff_ratios_640} displays how the cut-off energy follows a ratio of mass-to-charge to the power of 1.5--1.6, where the exponent is significantly smaller than the theoretical estimation of $2$.

%

\section{Discussion \& Conclusions}

Having studied the acceleration of multiple particle populations through self-generated turbulence with three different coronal shocks, we find that during early phases of the acceleration process, very hard, even flat spectra can be seen for high-mass ions. At all but the highest frequencies, the effect of minor ions on wave generation is non-negligible, especially in the region directly in front of the shock. As the shock-normal velocity increases, the deduced spectra become harder and the maximum energy attained increases. It is also seen that the  maximum energy dependence (Q/A)$^\beta$ does not exhibit $\beta=2$, as suggested by \citet{2007SSRv..130..255Z}. Rather, the behaviour of cut-off energies is between $\beta=1.5$ and $\beta=1.6$.

At high energies, accelerated particles stream away from the shock due to focusing, as scattering particles supply insufficient wave amplification power to trap high-energy particles to the shock. This causes the turbulent trapping boundary to approach and intersect the shock at the maximum ion energy. As the shock-normal velocity increases, the particle spectra become harder and the energy at which the turbulent trapping boundary intersects the shock increases.

To gauge the effect of focusing versus trapping on particle energy, we note that out of $V+D/L=V_\mathrm{s}$ only the diffusion coefficient $D\propto v^{2}/~\left(f_{\mathrm{ci}}f_{\mathrm{res}}P(f_{\mathrm{res}})\right)$ depends on the particle species. Thus, at the maximum energy edge of the trapping boundary, $D$ must be same for all particles. If at low frequencies the wave spectrum increases from ambient to amplified levels with a power law $P\propto f^{b}$, as shown in Fig. \ref{fig:wavespectra_both}, we find
\begin{equation}
D\propto\left(\frac{Q}{A}\right)^{-b-2}~ v^{b+3}.
\end{equation}
This gives the (non-relativistic) cutoff energy per nucleon a dependency of $(Q/A)^{2(b+2)/(b+3)}$. For a sharp cutoff, i.e. a purely rigidity-limited case, this results in a $(Q/A)^{2}$-dependence for the cutoff energy, while a smoother transition results in a significantly weaker dependence. Due to the weak dependence of $\beta$ on $b$, and the dynamic evolution of turbulence, care should be taken when inferring the turbulence spectrum shape from ion cutoff energies.

To examine particle acceleration dynamics, we can calculate the time $\tau_R$ required to accelerate a particle from an injection rigidity $R_0$ to a given rigidity $R=p/q_{i}\propto Av/Q$, where $R \gg R_0$. Assuming zero residence time in the downstream, we find 
\begin{equation}
 \tau_R = \mathcal{C} \int_{R_{0}}^{R} \frac{dR'}{R'} D(R') = \frac{Q}{A} \mathcal{C} \left( G(R) - G(R_0) \right),
\end{equation}
where $\mathcal{C}$ is a constant and $G(R)$ is a function of rigidity based on the shape of the wave power spectrum. If $P(f)\propto f^{b}$ where $b$ is constant, the acceleration time to a given rigidity $R$ is directly proportional to $Q/A$. Another item of interest is the time required to accelerate a heavy ion from injection speed $v_0$ to the maximum speed $v_\mathrm{ion}=(Q/A)^{(b+2)/(b+3)}v_\mathrm{p}$, where $v_p$ is the proton speed at the turbulent trapping boundary. This gives
\begin{align}
 \tau_\mathrm{trap} &= \mathcal{C} \int_{v_{0}}^{v_\mathrm{ion}} \left(\frac{Q}{A}\right)^{-b-2}(v')^{b+2} dv' \nonumber \\
 &\propto \left[ \left(\frac{Q}{A}\right)^{b+2}v_\mathrm{p}^{b+3} - v_0^{b+3} \right]\left(\frac{Q}{A}\right)^{-b-2}, \label{eq:trap acceleration time}
\end{align}
which, using previous assumptions for $b$, yields an acceleration time independent of the charge-to-mass ratio. Thus, it is clear that minor ions are accelerated to the maximum rigidity of protons much faster than the protons themselves, after which the ions slowly continue to gain energy until they reach the turbulent trapping boundary. This results in minor ions gaining harder, even flat spectra, especially in early phases of the simulation. In latter phases of the simulation, the value of $b$ increases at low frequencies, which leads to an increase in minor ion acceleration time and thus softer minor ion spectra.

At later stages of the simulations, wave amplification rates decay in line with the decay of injection efficiency. Thus, accelerated particles can stream away from the shock at lower energies, and further acceleration to higher energies ceases. Particles reaching the turbulent trapping boundary propagate in space, forming a plateau which is not consistent with Bell's steady-state result.

In conclusion, diffusive shock acceleration of protons and minor ions cannot realistically be represented by a steady-state approximation, but instead requires numerical simulations to reveal the full dynamics of the acceleration process and the various particle populations.

%

\begin{acknowledgements}
	The authors would like to thank the IT Center for Science Ltd (CSC) for computational
	services and the Academy of Finland (AF) for financial support of projects 122041 and 121650.
	TL acknowledges support from the UK Science and Technology Facilities
	Council (standard grant ST/H002944/1).
\end{acknowledgements}

%

\bibliographystyle{aa} 
\bibliography{17507_bib}

\end{document}